# ON THE ASYMPTOTIC SOLUTION FOR THE FOURIER-BESSEL MULTIPLE SCATTERING COEFFICIENTS OF AN INFINITE GRATING OF INSULATING DIELECTRIC CIRCULAR CYLINDERS AT OBLIQUE INCIDENCE

ÖMER KAVAKLIOĞLU and BARUCH SCHNEIDER

*Division of Electrophysics Research and Department of Mathematics,*

*Faculty of Computer Sciences, Izmir University of Economics,*

*Balçova, IZMIR 35330 TURKEY*

(omer_kavaklioglu@yahoo.com; omer.kavaklioglu@ieu.edu.tr; baruch.schneider@ieu.edu.tr)

**Abstract**

In this article, we present the *'asymptotic solution'* for the matrix system of equations representing the multiple scattering coefficients of an infinite grating of insulating dielectric circular cylinders associated with vertically polarized obliquely incident plane electromagnetic waves.

---

## 1. Introduction

Twersky [1-2] treated the classical electromagnetic problem of multiple scattering of waves by the infinite grating of dielectric circular cylinders at normal incidence and derived the equations describing the behavior of the multiple scattering coefficients of the infinite grating at normal incidence in terms of the *'Schlömilch series'* [3] and the *'scattering coefficients of an isolated cylinder at oblique incidence'* [4].

The exact equations representing the behavior of the *'Fourier-Bessel multiple scattering coefficients'* of an infinite grating of dielectric circular cylinders for obliquely incident and vertically polarized plane electromagnetic waves was first derived by Kavaklıoğlu [5-7]. The exact solution for these multiple scattering coefficients associated with the exterior electric and magnetic fields was acquired by the application of the *'direct Neumann iteration procedure'* in Kavaklıoğlu [8] in terms of *'Schlömilch series'* and the *'scattering coefficients of an isolated dielectric circular cylinder at oblique incidence'*, which was originally derived by Wait [4].

The purpose of this article is to acquire an asymptotic matrix representation for the *'Fourier-Bessel multiple scattering coefficients'* of the infinite grating corresponding to the vertically polarized and obliquely incident plane electromagnetic waves, and present the solution of the scattering coefficients as a function of *"the ratio of the cylinder radius to the grating spacing"* when the wavelength of the incident radiation is much greater than the grating spacing.

## 2. Description of the multiple scattering coefficients of the infinite grating at oblique incidence

In this section, we will present the exact systems of equations associated with the multiple scattering coefficients of the infinite grating of circular dielectric cylinders for obliquely incident and vertically polarized plane electromagnetic waves.

The exterior electric and magnetic fields of the infinite grating of circular dielectric cylinders, all having identical radii of '*a*', excited by obliquely incident and vertically polarized plane electromagnetic waves, are expressed in [5-7] in the coordinate system of the $s^{th}$ cylinder located at $\mathbf{r}_s$, in terms of the incident field plus a summation of cylindrical waves outgoing from each of the individual $m^{th}$ cylinder located at $\mathbf{r}_m$, for $|\mathbf{r}-\mathbf{r}_m| \to \infty$ as

$$E_z^{(ext)}(R_s,\phi_s,z) = \left\{ e^{ik_r sd \sin\psi_i} \sum_{n=-\infty}^{+\infty} \left[ \left( E_n^i + \sum_{m=-\infty}^{\infty} A_m \; I_{n-m}(k_r d) \right) J_n(k_r R_s) + A_n H_n^{(1)}(k_r R_s) \right] e^{in(\phi_s + \pi/2)} \right\} e^{-ik_z z} \tag{1a}$$

$$H_z^{(ext)}(R_s,\phi_s,z) = \left\{ e^{ik_r sd \sin\psi_i} \sum_{n=-\infty}^{+\infty} \left[ \left( \sum_{m=-\infty}^{\infty} A_m^H \; I_{n-m}(k_r d) \right) J_n(k_r R_s) + A_n^H H_n^{(1)}(k_r R_s) \right] e^{in(\phi_s + \pi/2)} \right\} e^{-ik_z z} \tag{1b}$$

In the representation of the exterior electric and magnetic fields above, the centers of the cylinders of the infinite grating are located at positions $\mathbf{r}_0$, $\mathbf{r}_1$, $\mathbf{r}_2$,.., etc., separated by a distance *'d'*, and $\{A_n, A_n^H\}_{n=-\infty}^{\infty}$ denotes the set of all *"multiple scattering Fourier-Bessel coefficients"* of the infinite grating corresponding to *"vertically polarized obliquely incident plane electromagnetic waves"*, associated with the exterior electric and magnetic fields respectively, $\forall n \ni n \in Z$, where "**Z**" represents the set of all integers. In the representation above, $\phi_i$ is the angle of incidence in *x-y* plane measured from $x-$axis in such a way that $\psi_i = \pi + \phi_i$ as it is indicated in figure 1, implying that the wave is arbitrarily incident in the first quadrant of the coordinate system and "$J_n(x)$" denotes Bessel function of order *'n'*. In expression (1a, b), we have

$$k_r = k_0 \sin\theta_i \tag{2a}$$

$$k_z = k_0 \cos\theta_i \tag{2b}$$

$k_0$ stands for the free space wave number with $k_0 := 2\pi / \lambda_0$, where $\lambda_0$ denotes the wavelength of the incident radiation, and $\theta_i$ is the *'obliquity angle'* made with $z-$axis. "$e^{-i\omega t}$" time dependence is suppressed throughout the article, where "$\omega$" represents the angular frequency of the incident wave in radians per second and "$t$"stands for time in seconds. In addition, we have

$$E_n^i = \sin\theta_i E_{0v} e^{-in\psi_i} \tag{3a}$$

$$\mathcal{I}_n(2\pi\Delta) = \sum_{p=1}^{+\infty} H_n^{(1)}(2\pi p\Delta)\left[e^{2\pi ip\Delta\sin\psi_i}(-1)^n + e^{-2\pi ip\Delta\sin\psi_i}\right] \tag{3b}$$

where $\Delta \equiv \frac{k_r d}{2\pi}$, and "$H_n^{(1)}(x)$" denotes the $n^{th}$ order Hankel function of first kind, The series in expression (3b) is the generalized form of the *'Schlömilch series [3] for obliquely incident waves* $\mathcal{I}_{n-m}(k_r d)$' [7] and convergent provided that $k_r d(1 \pm \sin\psi_i)/2\pi$ does not equal integers.

The multiple scattering coefficients associated with the exterior electric and magnetic fields of the infinite grating of dielectric circular cylinder corresponding to an obliquely incident vertically polarized plane wave have been acquired by the application of the separation-of-variables technique in Kavaklıoğlu [5] as

$$b_n^\mu \left\{ A_n + c_n \left[ E_n^i + \sum_{m=-\infty}^{+\infty} A_m \; \mathcal{I}_{n-m}(k_r d) \right] \right\} = -\left[ A_n^H + a_n^\mu \sum_{m=-\infty}^{+\infty} A_m^H \; \mathcal{I}_{n-m}(k_r d) \right] \tag{4a}$$

$\forall n \ni n \in Z$ , and

$$b_n^{\varepsilon}\left[\; A_n^H + c_n \sum_{m=-\infty}^{+\infty} A_m^H \;\; / \,_{n-m}(k_r d) \;\right] = A_n + a_n^{\varepsilon}\left[\; E_n^i + \sum_{m=-\infty}^{+\infty} A_m \;\; / \,_{n-m}(k_r d) \;\right] \tag{4b}$$

$\forall n \ni n \in Z$ . In these infinite set of equations, $A_n$ and $A_n^H$ represent the *'Fourier-Bessel multiple scattering coefficients'* corresponding to the electric and magnetic field intensities associated with obliquely incident vertically polarized plane electromagnetic waves, respectively. The coefficients appearing in the infinite set of linear algebraic equations above are defined as

$$c_n := \frac{J_n(k_r a)}{H_n^{(1)}(k_r a)} \tag{5}$$

$\forall n \ni n \in Z$ . Two sets of constants $a_n^{\zeta}$ and $b_n^{\zeta}$ appearing in the equations (4a, b), in which $\zeta_r \in \{\varepsilon_r, \mu_r\}$ stands for the dielectric constant and relative permeability of the dielectric cylinders respectively, are given as

$$a_n^{\zeta} = \left[ \frac{J_n(k_1 a)\dot{J}_n(k_r a) - \zeta_r\left(\dfrac{k_r}{k_1}\right)J_n(k_r a)\dot{J}_n(k_1 a)}{J_n(k_1 a)\dot{H}_n^{(1)}(k_r a) - \zeta_r\left(\dfrac{k_r}{k_1}\right)H_n^{(1)}(k_r a)\dot{J}_n(k_1 a)} \right] \tag{6}$$

$$b_n^{\zeta} = \sqrt{\frac{\varepsilon_0 \mu_0}{\zeta_0^2}} \left[ \frac{J_n(k_1 a)H_n^{(1)}(k_r a)}{J_n(k_1 a)\dot{H}_n^{(1)}(k_r a) - \zeta_r\left(\dfrac{k_r}{k_1}\right)H_n^{(1)}(k_r a)\dot{J}_n(k_1 a)} \right] \left(\frac{inF}{k_r a}\right) \tag{7}$$

for $\zeta \in \{\varepsilon, \mu\}$ , where $k_1$ is defined as $k_1 = k_0\sqrt{\varepsilon_r \mu_r - \cos^2\theta_i}$ , and *'F'* in the expression (7) above is given as

$$F = \frac{(\mu_r \varepsilon_r - 1)\cos\theta_i}{\mu_r \varepsilon_r - \cos^2\theta_i} \tag{8}$$

$\forall n \ni n \in Z$ . In these equations $\varepsilon_r$ and $\mu_r$ denotes the dielectric constant and the relative permeability constant of the insulating dielectric cylinders; $\varepsilon_0$ and $\mu_0$ stands for the permittivity and permeability of the free space respectively. The $\dot{J}_n$, and $\dot{H}_n^{(1)}$ in expressions (6-7) are defined as

$$\dot{J}_n(\varsigma) \equiv \tfrac{d}{d\varsigma} J_n(\varsigma) \tag{9a}$$

$$\dot{H}_n^{(1)}(\varsigma) \equiv \tfrac{d}{d\varsigma} H_n^{(1)}(\varsigma) \tag{9b}$$

which imply the first derivatives of the Bessel and Hankel functions of first kind and of order *n* with respect to their arguments.

## 3. Formulation of the asymptotic matrix equations for the scattering coefficients of the infinite grating at oblique incidence

In a previous investigation [9], the '*asymptotic equations associated with the multiple scattering coefficients of the electric and magnetic fields of an infinite grating of dielectric circular cylinders for an obliquely incident and vertically polarized plane wave'* was derived under the assumption that the '*wavelength of the incident radiation is much larger than the grating spacing'*. In this derivation, the behavior of the multiple scattering coefficients associated with obliquely incident and vertically polarized waves is estimated, when the wavelength of the incident field is much greater than the grating spacing, namely $(k_r a) << 1$ and $\left(\frac{k_r a}{k_r d}\right) \equiv \xi < \frac{1}{2}$, as

$$A_{\pm(2n-1)} \cong A_{\pm(2n-1),0} (k_r a)^{2n} \tag{10a}$$

$$A^{H}_{\pm(2n-1)} \cong A^{H}_{\pm(2n-1),0}(k_r a)^{2n} \tag{10b}$$

$\forall n \ni n \in N$, for the odd coefficients, and

$$A_{\pm 2n} \cong A_{\pm 2n,0}(k_r a)^{2n+2} \tag{10c}$$

$$A^{H}_{\pm 2n} \cong A^{H}_{\pm 2n,0}(k_r a)^{2n+2} \tag{10d}$$

$\forall n \ni n \in Z_+$, for the even coefficients. The conclusion declared above is achieved as a result of the detailed investigation of the behavior of the *'multiple scattering coefficients at oblique incidence'* and comparing their *'asymptotic behavior'* with those originally derived by Twersky [2] for the *'normal incidence'*. Defining a new $(2\times 1)$ vector $\underline{\varpi}_p$ for the $p^{th}$ multiple scattering coefficients of the infinite grating at oblique incidence as

$$\underline{\varpi}_p \equiv \begin{bmatrix} A_{p,0} \\ A^{H}_{p,0} \end{bmatrix}; \qquad \forall p \ni p \in Z \tag{11}$$

we have acquired the *'asymptotic matrix system of equations for the multiple scattering coefficients of the infinite grating of circular dielectric cylinders for obliquely incident and vertically polarized plane electromagnetic waves'* corresponding to *'odd'* subscripts as

$$
\begin{bmatrix} \bullet \\ \bullet \\ \underline{\varpi}_{+7} \\ \underline{\varpi}_{+5} \\ \underline{\varpi}_{+3} \\ \underline{\varpi}_{+1} \\ \underline{\varpi}_{-1} \\ \underline{\varpi}_{-3} \\ \underline{\varpi}_{-5} \\ \underline{\varpi}_{-7} \\ \bullet \\ \bullet \end{bmatrix}
=
\begin{bmatrix} \bullet \\ \bullet \\ \underline{0} \\ \underline{0} \\ \underline{0} \\ s_{+1}^{\varepsilon\mu} E_{+1}^{i} \\ s_{+1}^{\eta} E_{+1}^{i} \\ s_{+1}^{\varepsilon\mu} E_{-1}^{i} \\ s_{-1}^{\eta} E_{-1}^{i} \\ \underline{0} \\ \underline{0} \\ \underline{0} \\ \bullet \\ \bullet \end{bmatrix}
+
$$

$$
+\left[\begin{array}{cccc|cccc}
 & & & & \bullet & & \bullet & \bullet \\
 & & & & \left(\frac{a}{d}\right)^{8} h_{8} \underline{S}_{+7} & \left(\frac{a}{d}\right)^{10} h_{10} \underline{S}_{+7} & \left(\frac{a}{d}\right)^{12} h_{12} \underline{S}_{+7} & \left(\frac{a}{d}\right)^{14} h_{14} \underline{S}_{+7} \\
 & \underline{\mathbf{0}} & & & \left(\frac{a}{d}\right)^{6} h_{6} \underline{S}_{+5} & \left(\frac{a}{d}\right)^{8} h_{8} \underline{S}_{+5} & \left(\frac{a}{d}\right)^{10} h_{10} \underline{S}_{+5} & \left(\frac{a}{d}\right)^{12} h_{12} \underline{S}_{+5} \\
 & & & & \left(\frac{a}{d}\right)^{4} h_{4} \underline{S}_{+3} & \left(\frac{a}{d}\right)^{6} h_{6} \underline{S}_{+3} & \left(\frac{a}{d}\right)^{8} h_{8} \underline{S}_{+3} & \left(\frac{a}{d}\right)^{10} h_{10} \underline{S}_{+3} \\
 & & & & \left(\frac{a}{d}\right)^{2} h_{2} \underline{S}_{+1} & \left(\frac{a}{d}\right)^{4} h_{4} \underline{S}_{+1} & \left(\frac{a}{d}\right)^{6} h_{6} \underline{S}_{+1} & \left(\frac{a}{d}\right)^{8} h_{8} \underline{S}_{+1} \\
\hline
\left(\frac{a}{d}\right)^{8} h_{8} \underline{S}_{-1} & \left(\frac{a}{d}\right)^{6} h_{6} \underline{S}_{-1} & \left(\frac{a}{d}\right)^{4} h_{4} \underline{S}_{-1} & \left(\frac{a}{d}\right)^{2} h_{2} \underline{S}_{-1} & & & & \\
\left(\frac{a}{d}\right)^{10} h_{10} \underline{S}_{-3} & \left(\frac{a}{d}\right)^{8} h_{8} \underline{S}_{-3} & \left(\frac{a}{d}\right)^{6} h_{6} \underline{S}_{-3} & \left(\frac{a}{d}\right)^{4} h_{4} \underline{S}_{-3} & & & \underline{\mathbf{0}} & \\
\left(\frac{a}{d}\right)^{12} h_{12} \underline{S}_{-5} & \left(\frac{a}{d}\right)^{10} h_{10} \underline{S}_{-5} & \left(\frac{a}{d}\right)^{8} h_{8} \underline{S}_{-5} & \left(\frac{a}{d}\right)^{6} h_{6} \underline{S}_{-5} & & & & \\
\left(\frac{a}{d}\right)^{14} h_{14} \underline{S}_{-7} & \left(\frac{a}{d}\right)^{12} h_{12} \underline{S}_{-7} & \left(\frac{a}{d}\right)^{10} h_{10} \underline{S}_{-7} & \left(\frac{a}{d}\right)^{8} h_{8} \underline{S}_{-7} & & & & \\
\bullet & \bullet & \bullet & & & & &
\end{array}\right]
\begin{bmatrix} \bullet \\ \bullet \\ \underline{\varpi}_{+7} \\ \underline{\varpi}_{+5} \\ \underline{\varpi}_{+3} \\ \underline{\varpi}_{+1} \\ \underline{\varpi}_{-1} \\ \underline{\varpi}_{-3} \\ \underline{\varpi}_{-5} \\ \underline{\varpi}_{-7} \\ \bullet \\ \bullet \end{bmatrix}
\tag{12a}
$$

In the equation above, the unknown $(\infty \times 1)$ vector associated with the *'odd'* multiple scattering coefficients of the magnetic and electric field appears on both side of this equation. Replacing the unknown $(\infty \times 1)$ vector of the *'odd'* multiple scattering coefficients to the left hand side of (12a), we have obtained the matrix system of equations for the *'odd orders of multiple scattering coefficients of the infinite grating at oblique incidence'* as

$$\begin{bmatrix} -\underline{\mathrm{I}} & \begin{matrix} \bullet & & \bullet & & \bullet \\ \left(\frac{a}{d}\right)^{8} h_8 \underline{S}_{+7} & \left(\frac{a}{d}\right)^{10} h_{10} \underline{S}_{+7} & \left(\frac{a}{d}\right)^{12} h_{12} \underline{S}_{+7} & \left(\frac{a}{d}\right)^{14} h_{14} \underline{S}_{+7} & \bullet \\ \left(\frac{a}{d}\right)^{6} h_6 \underline{S}_{+5} & \left(\frac{a}{d}\right)^{8} h_8 \underline{S}_{+5} & \left(\frac{a}{d}\right)^{10} h_{10} \underline{S}_{+5} & \left(\frac{a}{d}\right)^{12} h_{12} \underline{S}_{+5} & \\ \left(\frac{a}{d}\right)^{4} h_4 \underline{S}_{+3} & \left(\frac{a}{d}\right)^{6} h_6 \underline{S}_{+3} & \left(\frac{a}{d}\right)^{8} h_8 \underline{S}_{+3} & \left(\frac{a}{d}\right)^{10} h_{10} \underline{S}_{+3} & \bullet \\ \left(\frac{a}{d}\right)^{2} h_2 \underline{S}_{+1} & \left(\frac{a}{d}\right)^{4} h_4 \underline{S}_{+1} & \left(\frac{a}{d}\right)^{6} h_6 \underline{S}_{+1} & \left(\frac{a}{d}\right)^{8} h_8 \underline{S}_{+1} & \end{matrix} \\ \begin{matrix} \bullet & \left(\frac{a}{d}\right)^{8} h_8 \underline{S}_{-1} & \left(\frac{a}{d}\right)^{6} h_6 \underline{S}_{-1} & \left(\frac{a}{d}\right)^{4} h_4 \underline{S}_{-1} & \left(\frac{a}{d}\right)^{2} h_2 \underline{S}_{-1} \\ & \left(\frac{a}{d}\right)^{10} h_{10} \underline{S}_{-3} & \left(\frac{a}{d}\right)^{8} h_8 \underline{S}_{-3} & \left(\frac{a}{d}\right)^{6} h_6 \underline{S}_{-3} & \left(\frac{a}{d}\right)^{4} h_4 \underline{S}_{-3} \\ \bullet & \left(\frac{a}{d}\right)^{12} h_{12} \underline{S}_{-5} & \left(\frac{a}{d}\right)^{10} h_{10} \underline{S}_{-5} & \left(\frac{a}{d}\right)^{8} h_8 \underline{S}_{-5} & \left(\frac{a}{d}\right)^{6} h_6 \underline{S}_{-5} \\ & \left(\frac{a}{d}\right)^{14} h_{14} \underline{S}_{-7} & \left(\frac{a}{d}\right)^{12} h_{12} \underline{S}_{-7} & \left(\frac{a}{d}\right)^{10} h_{10} \underline{S}_{-7} & \left(\frac{a}{d}\right)^{8} h_8 \underline{S}_{-7} \\ \bullet & & \bullet & \bullet & \end{matrix} & -\underline{\mathrm{I}} \end{bmatrix} \begin{bmatrix} \bullet \\ \bullet \\ \underline{\varpi}_{+7} \\ \underline{\varpi}_{+5} \\ \underline{\varpi}_{+3} \\ \underline{\varpi}_{+1} \\ \underline{\varpi}_{-1} \\ \underline{\varpi}_{-3} \\ \underline{\varpi}_{-5} \\ \underline{\varpi}_{-7} \\ \bullet \\ \bullet \end{bmatrix}$$

$$= -\begin{bmatrix} \bullet \\ \underline{0} \\ \underline{0} \\ \underline{0} \\ s_1^{\varepsilon\mu} E_{+1}^{i} \\ s_1^{\eta} E_{+1}^{i} \\ s_1^{\varepsilon\mu} E_{-1}^{i} \\ s_{-1}^{\eta} E_{-1}^{i} \\ \underline{0} \\ \underline{0} \\ \underline{0} \\ \bullet \end{bmatrix} \qquad (12b)$$

The *'asymptotic system matrix'* described by the equation (12b) above is of the order of $(\infty \times \infty)$, and the unknown is an $\infty-$dimensional vector corresponding to the *'odd orders'* of the multiple scattering coefficients of the infinite grating. Similarly, we have acquired the *'asymptotic matrix system of equations for the even orders of multiple scattering coefficients associated with the exterior electric and magnetic fields at oblique incidence'* as

$$
\begin{bmatrix} \bullet \\ \bullet \\ \underline{\varpi}_{+8} \\ \underline{\varpi}_{+6} \\ \underline{\varpi}_{+4} \\ \underline{\varpi}_{+2} \\ \underline{\varpi}_{-2} \\ \underline{\varpi}_{-4} \\ \underline{\varpi}_{-6} \\ \underline{\varpi}_{-8} \\ \bullet \\ \bullet \end{bmatrix}
=
\begin{bmatrix}
\underline{\mathbf{0}} &
\begin{matrix}
 & \bullet & & \bullet & \bullet \\
\left(\frac{a}{d}\right)^{10} h_{10} \underline{S}_{+8} & \left(\frac{a}{d}\right)^{12} h_{12} \underline{S}_{+8} & \left(\frac{a}{d}\right)^{14} h_{14} \underline{S}_{+8} & \left(\frac{a}{d}\right)^{16} h_{16} \underline{S}_{+8} & \bullet \\
\left(\frac{a}{d}\right)^{8} h_{8} \underline{S}_{+6} & \left(\frac{a}{d}\right)^{10} h_{10} \underline{S}_{+6} & \left(\frac{a}{d}\right)^{12} h_{12} \underline{S}_{+6} & \left(\frac{a}{d}\right)^{14} h_{14} \underline{S}_{+6} & \\
\left(\frac{a}{d}\right)^{6} h_{6} \underline{S}_{+4} & \left(\frac{a}{d}\right)^{8} h_{8} \underline{S}_{+4} & \left(\frac{a}{d}\right)^{10} h_{10} \underline{S}_{+4} & \left(\frac{a}{d}\right)^{12} h_{12} \underline{S}_{+4} & \bullet \\
\left(\frac{a}{d}\right)^{4} h_{4} \underline{S}_{+2} & \left(\frac{a}{d}\right)^{6} h_{6} \underline{S}_{+2} & \left(\frac{a}{d}\right)^{8} h_{8} \underline{S}_{+2} & \left(\frac{a}{d}\right)^{10} h_{10} \underline{S}_{+2} &
\end{matrix} \\
\begin{matrix}
\bullet & \left(\frac{a}{d}\right)^{10} h_{10} \underline{S}_{-2} & \left(\frac{a}{d}\right)^{8} h_{8} \underline{S}_{-2} & \left(\frac{a}{d}\right)^{6} h_{6} \underline{S}_{-2} & \left(\frac{a}{d}\right)^{4} h_{4} \underline{S}_{-2} \\
 & \left(\frac{a}{d}\right)^{12} h_{12} \underline{S}_{-4} & \left(\frac{a}{d}\right)^{10} h_{10} \underline{S}_{-4} & \left(\frac{a}{d}\right)^{8} h_{8} \underline{S}_{-4} & \left(\frac{a}{d}\right)^{6} h_{6} \underline{S}_{-4} \\
\bullet & \left(\frac{a}{d}\right)^{14} h_{14} \underline{S}_{-6} & \left(\frac{a}{d}\right)^{12} h_{12} \underline{S}_{-6} & \left(\frac{a}{d}\right)^{10} h_{10} \underline{S}_{-6} & \left(\frac{a}{d}\right)^{8} h_{8} \underline{S}_{-6} \\
 & \left(\frac{a}{d}\right)^{16} h_{16} \underline{S}_{-8} & \left(\frac{a}{d}\right)^{14} h_{14} \underline{S}_{-8} & \left(\frac{a}{d}\right)^{12} h_{12} \underline{S}_{-8} & \left(\frac{a}{d}\right)^{10} h_{10} \underline{S}_{-8} \\
\bullet & & \bullet & & \bullet
\end{matrix} & \underline{\mathbf{0}}
\end{bmatrix}
\begin{bmatrix} \bullet \\ \bullet \\ \underline{\varpi}_{+8} \\ \underline{\varpi}_{+6} \\ \underline{\varpi}_{+4} \\ \underline{\varpi}_{+2} \\ \underline{\varpi}_{-2} \\ \underline{\varpi}_{-4} \\ \underline{\varpi}_{-6} \\ \underline{\varpi}_{-8} \\ \bullet \\ \bullet \end{bmatrix}
$$

$$
+
\begin{bmatrix}
\underline{\mathbf{0}} &
\begin{matrix}
 & \bullet & & \bullet & \bullet \\
\left(\frac{a}{d}\right)^{8} h_{+9} \underline{S}_{+8} & \left(\frac{a}{d}\right)^{10} h_{+11} \underline{S}_{+8} & \left(\frac{a}{d}\right)^{12} h_{+13} \underline{S}_{+8} & \left(\frac{a}{d}\right)^{14} h_{+15} \underline{S}_{+8} & \bullet \\
\left(\frac{a}{d}\right)^{6} h_{+7} \underline{S}_{+6} & \left(\frac{a}{d}\right)^{8} h_{+9} \underline{S}_{+6} & \left(\frac{a}{d}\right)^{10} h_{+11} \underline{S}_{+6} & \left(\frac{a}{d}\right)^{12} h_{+13} \underline{S}_{+6} & \\
\left(\frac{a}{d}\right)^{4} h_{+5} \underline{S}_{+4} & \left(\frac{a}{d}\right)^{6} h_{+7} \underline{S}_{+4} & \left(\frac{a}{d}\right)^{8} h_{+9} \underline{S}_{+4} & \left(\frac{a}{d}\right)^{10} h_{+11} \underline{S}_{+4} & \bullet \\
\left(\frac{a}{d}\right)^{2} h_{+3} \underline{S}_{+2} & \left(\frac{a}{d}\right)^{4} h_{+5} \underline{S}_{+2} & \left(\frac{a}{d}\right)^{6} h_{+7} \underline{S}_{+2} & \left(\frac{a}{d}\right)^{8} h_{+9} \underline{S}_{+2} &
\end{matrix} \\
\begin{matrix}
\bullet & \left(\frac{a}{d}\right)^{8} h_{-9} \underline{S}_{-2} & \left(\frac{a}{d}\right)^{6} h_{-7} \underline{S}_{-2} & \left(\frac{a}{d}\right)^{4} h_{-5} \underline{S}_{-2} & \left(\frac{a}{d}\right)^{2} h_{-3} \underline{S}_{-2} \\
 & \left(\frac{a}{d}\right)^{10} h_{-11} \underline{S}_{-4} & \left(\frac{a}{d}\right)^{8} h_{-9} \underline{S}_{-4} & \left(\frac{a}{d}\right)^{6} h_{-7} \underline{S}_{-4} & \left(\frac{a}{d}\right)^{4} h_{-5} \underline{S}_{-4} \\
\bullet & \left(\frac{a}{d}\right)^{12} h_{-13} \underline{S}_{-6} & \left(\frac{a}{d}\right)^{10} h_{-11} \underline{S}_{-6} & \left(\frac{a}{d}\right)^{8} h_{-9} \underline{S}_{-6} & \left(\frac{a}{d}\right)^{6} h_{-7} \underline{S}_{-6} \\
 & \left(\frac{a}{d}\right)^{14} h_{-15} \underline{S}_{-8} & \left(\frac{a}{d}\right)^{12} h_{-13} \underline{S}_{-8} & \left(\frac{a}{d}\right)^{10} h_{-11} \underline{S}_{-8} & \left(\frac{a}{d}\right)^{8} h_{-9} \underline{S}_{-8} \\
\bullet & & \bullet & & \bullet
\end{matrix} & \underline{\mathbf{0}}
\end{bmatrix}
\begin{bmatrix} \bullet \\ \bullet \\ \underline{\varpi}_{+7} \\ \underline{\varpi}_{+5} \\ \underline{\varpi}_{+3} \\ \underline{\varpi}_{+1} \\ \underline{\varpi}_{-1} \\ \underline{\varpi}_{-3} \\ \underline{\varpi}_{-5} \\ \underline{\varpi}_{-7} \\ \bullet \\ \bullet \end{bmatrix}
$$

$$
+\begin{bmatrix}
\left[\left(\frac{a}{d}\right)^8 h_8 s_0^{\varepsilon\mu} E_0^i\right]\begin{pmatrix} s_8^{\varepsilon\mu} \\ s_8^{\eta} \end{pmatrix} \\
\left[\left(\frac{a}{d}\right)^6 h_6 s_0^{\varepsilon\mu} E_0^i\right]\begin{pmatrix} s_6^{\varepsilon\mu} \\ s_6^{\eta} \end{pmatrix} \\
\left[\left(\frac{a}{d}\right)^4 h_4 s_0^{\varepsilon\mu} E_0^i\right]\begin{pmatrix} s_4^{\varepsilon\mu} \\ s_4^{\eta} \end{pmatrix} \\
\left[E_{+2}^i + \left(\frac{a}{d}\right)^2 h_2 s_0^{\varepsilon\mu} E_0^i\right]\begin{pmatrix} s_2^{\varepsilon\mu} \\ s_2^{\eta} \end{pmatrix} \\
\left[E_{-2}^i + \left(\frac{a}{d}\right)^2 h_2 s_0^{\varepsilon\mu} E_0^i\right]\begin{pmatrix} s_2^{\varepsilon\mu} \\ s_{-2}^{\eta} \end{pmatrix} \\
\left[\left(\frac{a}{d}\right)^4 h_4 s_0^{\varepsilon\mu} E_0^i\right]\begin{pmatrix} s_4^{\varepsilon\mu} \\ s_{-4}^{\eta} \end{pmatrix} \\
\left[\left(\frac{a}{d}\right)^6 h_6 s_0^{\varepsilon\mu} E_0^i\right]\begin{pmatrix} s_6^{\varepsilon\mu} \\ s_{-6}^{\eta} \end{pmatrix} \\
\left[\left(\frac{a}{d}\right)^8 h_8 s_0^{\varepsilon\mu} E_0^i\right]\begin{pmatrix} s_8^{\varepsilon\mu} \\ s_{-8}^{\eta} \end{pmatrix}
\end{bmatrix} \tag{13a}
$$

In the equation above, the unknown $\infty-$dimensional vector for the *'even orders'* of the multiple scattering coefficients associated with the electric and magnetic fields at oblique incidence appears on both side of this equation, and the solution from the equation (12b) for the *'odd multiple scattering coefficients'* appears in the *'asymptotic system of equations for even orders of multiple scattering coefficients'* in (13a) as known values. Taking the unknown $\infty-$dimensional vector for the *'even multiple scattering coefficients'* to the left hand side of (13a), we have obtained the *'asymptotic form of the matrix system of equations for the even orders of multiple scattering coefficients at oblique incidence,'* when the grating spacing is small compare to a wavelength, i.e., $(k_r d) << 1$, and $\left(\frac{k_r a}{k_r d}\right) \equiv \xi < \frac{1}{2}$, as

$$\left[\begin{array}{cc} -\underline{\mathbf{I}} & \begin{matrix} \left(\frac{a}{d}\right)^{10} h_{10}\underline{S}_{+8} & \left(\frac{a}{d}\right)^{12} h_{12}\underline{S}_{+8} & \left(\frac{a}{d}\right)^{14} h_{14}\underline{S}_{+8} & \left(\frac{a}{d}\right)^{16} h_{16}\underline{S}_{+8} \\ \left(\frac{a}{d}\right)^{8} h_{8}\underline{S}_{+6} & \left(\frac{a}{d}\right)^{10} h_{10}\underline{S}_{+6} & \left(\frac{a}{d}\right)^{12} h_{12}\underline{S}_{+6} & \left(\frac{a}{d}\right)^{14} h_{14}\underline{S}_{+6} \\ \left(\frac{a}{d}\right)^{6} h_{6}\underline{S}_{+4} & \left(\frac{a}{d}\right)^{8} h_{8}\underline{S}_{+4} & \left(\frac{a}{d}\right)^{10} h_{10}\underline{S}_{+4} & \left(\frac{a}{d}\right)^{12} h_{12}\underline{S}_{+4} \\ \left(\frac{a}{d}\right)^{4} h_{4}\underline{S}_{+2} & \left(\frac{a}{d}\right)^{6} h_{6}\underline{S}_{+2} & \left(\frac{a}{d}\right)^{8} h_{8}\underline{S}_{+2} & \left(\frac{a}{d}\right)^{10} h_{10}\underline{S}_{+2} \end{matrix} \\ \begin{matrix} \left(\frac{a}{d}\right)^{10} h_{10}\underline{S}_{-2} & \left(\frac{a}{d}\right)^{8} h_{8}\underline{S}_{-2} & \left(\frac{a}{d}\right)^{6} h_{6}\underline{S}_{-2} & \left(\frac{a}{d}\right)^{4} h_{4}\underline{S}_{-2} \\ \left(\frac{a}{d}\right)^{12} h_{12}\underline{S}_{-4} & \left(\frac{a}{d}\right)^{10} h_{10}\underline{S}_{-4} & \left(\frac{a}{d}\right)^{8} h_{8}\underline{S}_{-4} & \left(\frac{a}{d}\right)^{6} h_{6}\underline{S}_{-4} \\ \left(\frac{a}{d}\right)^{14} h_{14}\underline{S}_{-6} & \left(\frac{a}{d}\right)^{12} h_{12}\underline{S}_{-6} & \left(\frac{a}{d}\right)^{10} h_{10}\underline{S}_{-6} & \left(\frac{a}{d}\right)^{8} h_{8}\underline{S}_{-6} \\ \left(\frac{a}{d}\right)^{16} h_{16}\underline{S}_{-8} & \left(\frac{a}{d}\right)^{14} h_{14}\underline{S}_{-8} & \left(\frac{a}{d}\right)^{12} h_{12}\underline{S}_{-8} & \left(\frac{a}{d}\right)^{10} h_{10}\underline{S}_{-8} \end{matrix} & -\underline{\mathbf{I}} \end{array}\right] \begin{bmatrix} \bullet \\ \bullet \\ \underline{\varpi}_{+8} \\ \underline{\varpi}_{+6} \\ \underline{\varpi}_{+4} \\ \underline{\varpi}_{+2} \\ \underline{\varpi}_{-2} \\ \underline{\varpi}_{-4} \\ \underline{\varpi}_{-6} \\ \underline{\varpi}_{-8} \\ \bullet \\ \bullet \end{bmatrix} =$$

$$-\left[\begin{array}{cc} \underline{\mathbf{0}} & \begin{matrix} \left(\frac{a}{d}\right)^{8} h_{+9}\underline{S}_{+8} & \left(\frac{a}{d}\right)^{10} h_{+11}\underline{S}_{+8} & \left(\frac{a}{d}\right)^{12} h_{+13}\underline{S}_{+8} & \left(\frac{a}{d}\right)^{14} h_{+15}\underline{S}_{+8} \\ \left(\frac{a}{d}\right)^{6} h_{+7}\underline{S}_{+6} & \left(\frac{a}{d}\right)^{8} h_{+9}\underline{S}_{+6} & \left(\frac{a}{d}\right)^{10} h_{+11}\underline{S}_{+6} & \left(\frac{a}{d}\right)^{12} h_{+13}\underline{S}_{+6} \\ \left(\frac{a}{d}\right)^{4} h_{+5}\underline{S}_{+4} & \left(\frac{a}{d}\right)^{6} h_{+7}\underline{S}_{+4} & \left(\frac{a}{d}\right)^{8} h_{+9}\underline{S}_{+4} & \left(\frac{a}{d}\right)^{10} h_{+11}\underline{S}_{+4} \\ \left(\frac{a}{d}\right)^{2} h_{+3}\underline{S}_{+2} & \left(\frac{a}{d}\right)^{4} h_{+5}\underline{S}_{+2} & \left(\frac{a}{d}\right)^{6} h_{+7}\underline{S}_{+2} & \left(\frac{a}{d}\right)^{8} h_{+9}\underline{S}_{+2} \end{matrix} \\ \begin{matrix} \left(\frac{a}{d}\right)^{8} h_{-9}\underline{S}_{-2} & \left(\frac{a}{d}\right)^{6} h_{-7}\underline{S}_{-2} & \left(\frac{a}{d}\right)^{4} h_{-5}\underline{S}_{-2} & \left(\frac{a}{d}\right)^{2} h_{-3}\underline{S}_{-2} \\ \left(\frac{a}{d}\right)^{10} h_{-11}\underline{S}_{-4} & \left(\frac{a}{d}\right)^{8} h_{-9}\underline{S}_{-4} & \left(\frac{a}{d}\right)^{6} h_{-7}\underline{S}_{-4} & \left(\frac{a}{d}\right)^{4} h_{-5}\underline{S}_{-4} \\ \left(\frac{a}{d}\right)^{12} h_{-13}\underline{S}_{-6} & \left(\frac{a}{d}\right)^{10} h_{-11}\underline{S}_{-6} & \left(\frac{a}{d}\right)^{8} h_{-9}\underline{S}_{-6} & \left(\frac{a}{d}\right)^{6} h_{-7}\underline{S}_{-6} \\ \left(\frac{a}{d}\right)^{14} h_{-15}\underline{S}_{-8} & \left(\frac{a}{d}\right)^{12} h_{-13}\underline{S}_{-8} & \left(\frac{a}{d}\right)^{10} h_{-11}\underline{S}_{-8} & \left(\frac{a}{d}\right)^{8} h_{-9}\underline{S}_{-8} \end{matrix} & \underline{\mathbf{0}} \end{array}\right] \begin{bmatrix} \bullet \\ \bullet \\ \underline{\varpi}_{+7} \\ \underline{\varpi}_{+5} \\ \underline{\varpi}_{+3} \\ \underline{\varpi}_{+1} \\ \underline{\varpi}_{-1} \\ \underline{\varpi}_{-3} \\ \underline{\varpi}_{-5} \\ \underline{\varpi}_{-7} \\ \bullet \\ \bullet \end{bmatrix}$$

$$
+\begin{bmatrix}
\left[\left(\tfrac{a}{d}\right)^8 h_8 s_0^{\varepsilon\mu} E_0^i\right]\begin{pmatrix} s_8^{\varepsilon\mu} \\ s_8^{\eta} \end{pmatrix} \\
\left[\left(\tfrac{a}{d}\right)^6 h_6 s_0^{\varepsilon\mu} E_0^i\right]\begin{pmatrix} s_6^{\varepsilon\mu} \\ s_6^{\eta} \end{pmatrix} \\
\left[\left(\tfrac{a}{d}\right)^4 h_4 s_0^{\varepsilon\mu} E_0^i\right]\begin{pmatrix} s_4^{\varepsilon\mu} \\ s_4^{\eta} \end{pmatrix} \\
\left[E_{+2}^i + \left(\tfrac{a}{d}\right)^2 h_2 s_0^{\varepsilon\mu} E_0^i\right]\begin{pmatrix} s_2^{\varepsilon\mu} \\ s_2^{\eta} \end{pmatrix} \\
\left[E_{-2}^i + \left(\tfrac{a}{d}\right)^2 h_2 s_0^{\varepsilon\mu} E_0^i\right]\begin{pmatrix} s_2^{\varepsilon\mu} \\ s_{-2}^{\eta} \end{pmatrix} \\
\left[\left(\tfrac{a}{d}\right)^4 h_4 s_0^{\varepsilon\mu} E_0^i\right]\begin{pmatrix} s_4^{\varepsilon\mu} \\ s_{-4}^{\eta} \end{pmatrix} \\
\left[\left(\tfrac{a}{d}\right)^6 h_6 s_0^{\varepsilon\mu} E_0^i\right]\begin{pmatrix} s_6^{\varepsilon\mu} \\ s_{-6}^{\eta} \end{pmatrix} \\
\left[\left(\tfrac{a}{d}\right)^8 h_8 s_0^{\varepsilon\mu} E_0^i\right]\begin{pmatrix} s_8^{\varepsilon\mu} \\ s_{-8}^{\eta} \end{pmatrix}
\end{bmatrix} \tag{13b}
$$

The *'asymptotic system matrix'* in equation (13b) is of the order of $(\infty \times \infty)$ and the unknown is an $\infty-$dimensional vector corresponding to the *'even orders'* of the multiple scattering coefficients of the electric and magnetic fields associated with vertically polarized obliquely incident waves. In the equations (12 an 13), $\underline{\underline{S}}_n$ is a $(2\times 2)$ matrix defined as

$$
\underline{\underline{S}}_{\pm n} := \frac{1}{D}\begin{pmatrix} s_n^{\varepsilon\mu} & s_{\pm n}^{\xi} \\ s_{\pm n}^{\eta} & s_n^{\mu\varepsilon} \end{pmatrix}(k_r a)^{2n} \tag{14}
$$

In the above, we have

$$D = \left[1 + \varepsilon_r \left(\frac{k_r}{k_1}\right)^2\right]\left[1 + \mu_r \left(\frac{k_r}{k_1}\right)^2\right] - F^2 \tag{15}$$

The $n$-dependent constants appearing in the matrix of expression (14) are defined as

$$s_n^{\varepsilon\mu} := \left[\frac{in\pi}{(2^n n!)^2}\right] s_{\varepsilon\mu} \tag{16a}$$

$$s_{\pm n}^{\xi} := \left[\frac{in\pi}{(2^n n!)^2}\right] s_{\pm\xi} \tag{16b}$$

$$s_{\pm n}^{\eta} := \left[\frac{in\pi}{(2^n n!)^2}\right] s_{\pm\eta} \tag{16c}$$

$$s_n^{\mu\varepsilon} := \left[\frac{in\pi}{(2^n n!)^2}\right] s_{\mu\varepsilon} \tag{16d}$$

$\forall n \ni n \in N$ , where '$N$' denotes the set of all natural numbers. The various constants appearing in the definitions of (16) are expressed as

$$s_{\varepsilon\mu} = \left[1 - \varepsilon_r \left(\frac{k_r}{k_1}\right)^2\right]\left[1 + \mu_r \left(\frac{k_r}{k_1}\right)^2\right] + F^2 \tag{17a}$$

$$s_{\mu\varepsilon} = \left[1 - \mu_r \left(\frac{k_r}{k_1}\right)^2\right]\left[1 + \varepsilon_r \left(\frac{k_r}{k_1}\right)^2\right] + F^2 \tag{17b}$$

$$s_{\pm\xi} = \pm 2i\xi_0 F \tag{17c}$$

$$s_{\pm\eta} = \mp 2i\eta_0 F \tag{17d}$$

In addition, we have recognized that the matrix $\underline{\underline{s}}_n$ in equations (12-13) can be rewritten using (16) as

$$\underline{\underline{S}}_{\pm n} := \frac{1}{D}\left(\frac{in\pi}{(2^n n!)^2}\right)\begin{pmatrix} s_{\varepsilon\mu} & s_{\pm\xi} \\ s_{\pm\eta} & s_{\mu\varepsilon} \end{pmatrix}(k_r a)^{2n} \tag{18}$$

The $h_n$'s arising in the asymptotic equations of (12-13) represents the leading asymptotic terms of the *'Schlömilch Series'*, which are expressed in ref. [2-3] for normal incidence, and in ref. [7] for oblique incidence as

$$\mathsf{H}_0 \approx \frac{h_0}{k_r d} \qquad \text{where} \qquad h_0 \equiv 2\sec\phi_0 \tag{19a}$$

$$\mathsf{H}_1 \approx \frac{h_1}{k_r d} \qquad \text{where} \qquad h_1 \equiv -2i\tan\phi_0 \tag{19b}$$

$$\mathsf{H}_2 \approx \frac{h_2}{(k_r d)^2} \qquad \text{where} \qquad h_2 \equiv \frac{4\pi}{3i} \tag{19c}$$

$$\mathsf{H}_3 \approx \frac{h_3}{(k_r d)^2} \qquad \text{where} \qquad h_3 \equiv -\frac{16\pi\sin\phi_0}{3} \tag{19d}$$

$$\mathsf{H}_4 \approx \frac{h_4}{(k_r d)^4} \qquad \text{where} \qquad h_4 \equiv \frac{2^5\pi^3}{15i} \tag{19e}$$

$$\mathsf{H}_5 \approx \frac{h_5}{(k_r d)^4} \qquad \text{where} \qquad h_5 \equiv -\frac{2^8\pi^3\sin\phi_0}{15} \tag{19f}$$

The leading terms of $\mathsf{H}_n$ for large $n$ in general are given by

$$\mathsf{H}_{2n} \approx \frac{h_{2n}}{(k_r d)^{2n}} \tag{20a}$$

$$\mathsf{H}_{2n+1} \approx \frac{h_{2n+1}}{(k_r d)^{2n}} \tag{20b}$$

where $h_{2n}$'s and $h_{2n+1}$'s for large $n$ are given as

$$h_{2n} \to \frac{i}{n}(-1)^n 2^{4n-1}\pi^{2n-1}B_{2n}(0) \tag{21a}$$

and

$$h_{2n+1} \to (-1)^n 2^{4n+1}\pi^{2n-1}B_{2n}(0)\sin\phi_0 \equiv -4inh_{2n}\sin\phi_0 \tag{21b}$$

respectively. In the expressions above, $B_\xi$'s are the *'Bernoulli numbers',* and the relationship between *'Bernoulli polynomial'* and *'Bernoulli numbers'* is given as $B_{2\xi}(0) \equiv (-1)^{\xi-1} B_\xi$ .

**4. Solution for the matrix system of equations for the scattering coefficients of the infinite grating**

The solution for the *'asymptotic matrix system'* of equations, given in (12b) and (13b), can then be obtained by expanding '$\underline{\varpi}_n$' in the form of an *'infinite asymptotic series expansion'*. For this purpose, we have introduced

$$\underline{\varpi}_n = \sum_{m=0}^{\infty} \underline{\varpi}_n^{(m)} \left( \frac{a}{d} \right)^m \tag{22}$$

$\forall n \ni n \in N$ , into (12b, 13b) and determined the coefficients in the *'asymptotic series expansion'* in (22), namely, $\underline{\varpi}_n^{(m)}$ for *"m=0, 1, 2, 3,.."*.

The multiple scattering coefficients corresponding to the fundamental mode are given by

$$A_{0,0} \cong \sin\theta_i s_0^{\varepsilon\mu} \tag{23a}$$

for the scattered electric field at oblique incidence, and

$$A_{0,0}^H \equiv 0 \tag{23b}$$

for the scattered magnetic field at oblique incidence, where $s_0^{\varepsilon\mu}$ in (23a) is given by

$$s_0^{\varepsilon\mu} = \frac{i\pi}{4}(\varepsilon_r - 1) \tag{23c}$$

We have obtained for $n = 1$

$$A_{\pm1,0} \cong \sin\theta_i\left(\frac{i\pi}{4D}\right)\left\{s_{\varepsilon\mu}e^{\mp i\psi_i} + \left(\frac{a}{d}\right)^2 h_2(s_{\varepsilon\mu}^2 - 4F^2)\left(\frac{i\pi}{4D}\right)e^{\pm i\psi_i}\right.$$

$$\left. + \left(\frac{a}{d}\right)^4 h_2^2\left[s_{\varepsilon\mu}(s_{\varepsilon\mu}^2 - 4F^2) + 8F^2(\varepsilon_r - \mu_r)\left(\frac{k_r}{k_1}\right)^2\right]\left(\frac{i\pi}{4D}\right)^2 e^{\mp i\psi_i} + O\left(\left(\frac{a}{d}\right)^6\right)\right\} \quad (24a)$$

for the scattered electric field at oblique incidence, and

$$A_{\pm1,0}^H \cong \mp 2i\eta_0 F\sin\theta_i\left(\frac{i\pi}{4D}\right)\left\{e^{\mp i\psi_i} + \left(\frac{a}{d}\right)^2 h_2 2(\mu_r - \varepsilon_r)\left(\frac{k_r}{k_1}\right)^2\left(\frac{i\pi}{4D}\right)e^{\pm i\psi_i}\right.$$

$$\left. + \left(\frac{a}{d}\right)^4 h_2^2\left[(s_{\mu\varepsilon}^2 - 4F^2) + 2(\mu_r - \varepsilon_r)\left(\frac{k_r}{k_1}\right)^2 s_{\varepsilon\mu}\right]\left(\frac{i\pi}{4D}\right)^2 e^{\mp i\psi_i} + O\left(\left(\frac{a}{d}\right)^6\right)\right\} \quad (24b)$$

for the scattered magnetic field at oblique incidence, respectively. Similarly, for $n = 2$, we have obtained

$$A_{\pm2,0} \cong \sin\theta_i\left(\frac{i\pi}{32D}\right)\left\{s_{\varepsilon\mu}e^{\mp 2i\psi_i} + \left(\frac{a}{d}\right)^2\left(h_2 s_0^{\varepsilon\mu}s_{\varepsilon\mu} \pm h_3(s_{\varepsilon\mu}^2 - 4F^2)\left(\frac{i\pi}{4D}\right)e^{\pm i\psi_i}\right)\right.$$

$$+ \left(\frac{a}{d}\right)^4\left[h_4\left(\frac{i\pi}{32D}\right)(s_{\varepsilon\mu}^2 - 4F^2)e^{\pm 2i\psi_i}\right.$$

$$\left.\left. \pm h_5 h_2\left(\frac{i\pi}{4D}\right)^2\left(s_{\varepsilon\mu}(s_{\varepsilon\mu}^2 - 4F^2) + 8F^2(\varepsilon_r - \mu_r)\left(\frac{k_r}{k_1}\right)^2\right)e^{\mp i\psi_i}\right] + O\left(\left(\frac{a}{d}\right)^6\right)\right\} \quad (25a)$$

for the scattered electric field at oblique incidence, and

$$A_{\pm2,0}^H \cong \mp 2i\eta_0 F\sin\theta_i\left(\frac{i\pi}{32D}\right)$$

$$\cdot\left\{e^{\mp 2i\psi_i} + \left(\frac{a}{d}\right)^2\left(h_2 s_0^{\varepsilon\mu} \pm h_3\left(\frac{i\pi}{4D}\right)2(\mu_r - \varepsilon_r)\left(\frac{k_r}{k_1}\right)^2 e^{\pm i\psi_i}\right)\right.$$

$$+\left(\frac{a}{d}\right)^4\Bigg[h_4\left(\frac{i\pi}{32D}\right)2(\mu_r-\varepsilon_r)\left(\frac{k_r}{k_1}\right)^2 e^{\pm 2i\psi_i}$$

$$\pm h_5 h_2\left(\frac{i\pi}{4D}\right)^2\left((s_{\varepsilon\mu}^2-4F^2)+2(\varepsilon_r-\mu_r)\left(\frac{k_r}{k_1}\right)^2 s_{\mu\varepsilon}\right)e^{\mp i\psi_i}\Bigg]+O\left(\left(\frac{a}{d}\right)^6\right)\Bigg\} \quad (25b)$$

for the scattered magnetic field at oblique incidence, respectively. Finally, for $n=3$, we have acquired

$$A_{\pm 3,0}\cong \sin\theta_i\left(\frac{i\pi}{3.2^8 D}\right)\left[\left(\frac{a}{d}\right)^4 h_4(s_{\varepsilon\mu}^2-4F^2)\left(\frac{i\pi}{4D}\right)e^{\pm i\psi_i}+O\left(\left(\frac{a}{d}\right)^6\right)\right] \quad (26a)$$

for the scattered electric field at oblique incidence, and

$$A_{\pm 3,0}^H\cong \mp 2i\eta_0 F\sin\theta_i\left(\frac{i\pi}{3.2^8 D}\right)\left[\left(\frac{a}{d}\right)^4 h_4 2(\mu_r-\varepsilon_r)\left(\frac{k_r}{k_1}\right)^2\left(\frac{i\pi}{4D}\right)e^{\pm i\psi_i}+O\left(\left(\frac{a}{d}\right)^6\right)\right] \quad (26b)$$

for the scattered magnetic field at oblique incidence, respectively.

## 5. Conclusion

We have presented the *'asymptotic matrix system of equations of the multiple scattering coefficients'* of an infinite grating of circular dielectric cylinders for obliquely incident and vertically polarized plane electromagnetic waves associated with the exterior electric and magnetic field intensities. Furthermore, we have acquired the asymptotic solution for the multiple scattering coefficients up to and including third order as a function of the cylinder radius to grating spacing when the grating spacing *'d'* is small compare to a wavelength, i.e., $(k_r d) << 1$, and $\left(\frac{k_r a}{k_r d}\right)\equiv \xi < \frac{1}{2}$.

## Acknowledgments

We are indebted to deceased Professor Emeritus Victor Twersky of the Department of Mathematics of the University of Illinois at Chicago for the kind concern he spared to our work. The first author takes this opportunity to express his sincere thanks to Professor Dr Roger Henry Lang for suggesting the problem and many fruitful discussions.

## References

[1] Twersky, V. 1956 On the scattering of waves by an infinite grating. *IRE Trans. on Antennas Propagat.* **AP-4**, 330-345.

[2] Twersky, V. 1962 On scattering of waves by the infinite grating of circular cylinders. *IRE Trans. on Antennas Propagat.* **AP-10**, 737-765.

[3] Twersky, V. 1961 Elementary function representations of Schlömilch series. *Arch. for Rational Mech. and Anal.* **8**, 323-332.

[4] Wait, J. R. 1955 Scattering of a plane wave from a circular dielectric cylinder at oblique incidence. *Canad. J. Phys.* **33**, 189-195.

[5] Kavaklıoğlu, Ö. 2000 Scattering of a plane wave by an infinite grating of circular dielectric cylinders at oblique incidence: E-polarization. *Int. J. Electron.* 87, 315-336.

[6] Kavaklıoğlu, Ö. 2001 On diffraction of waves by the infinite grating of circular dielectric cylinders at oblique incidence: Floquet representation. *J. Mod. Opt.* **48**, 125-142.

[7] Kavaklıoğlu, Ö. 2002 On Schlömilch series representation for the transverse electric multiple scattering by an infinite grating of insulating dielectric circular cylinders at oblique incidence. *J. Phys. A: Math. Gen.* **35**, 2229-2248.

[8] Kavaklıoğlu, Ö. and Schneider, B. 2007 On multiple scattering of radiation by an infinite grating of dielectric circular cylinders at oblique incidence. *preprint/* *www.arXiv.org* /arXiv:0710.1602v1

[9] Kavaklıoğlu, Ö. and Schneider, B. 2007 On the asymptotic representation for transverse magnetic multiple scattering of radiation by an infinite grating of dielectric circular cylinders at oblique incidence (in review)